\documentclass[amsmath,amssymb,twocolumn,prc,showpacs,nofootinbib,floatfix,superscriptaddress]{revtex4-1}

\usepackage{breakcites}
\usepackage{float}
\usepackage{soul}
\usepackage{amsmath}
\usepackage{graphicx}
\usepackage{dcolumn}
\usepackage{bm}
\usepackage{natbib}
\usepackage{tikz}
\usepackage{color}

\usepackage{subfigure}
\usepackage{longtable}
\usepackage{dcolumn}

\usepackage[colorlinks, linkcolor={blue},citecolor={blue},breaklinks]{hyperref}

\begin{document}

\preprint{APS/123-QED}
\title{Superallowed $0^+ \rightarrow 0^+$ $\beta$ decay of $T =2$ $^{20}$Mg: $Q_{\textrm{EC}}$ value and $\beta\gamma$ branching}

\author{B.~E.~Glassman}
\email{brenteglassman@gmail.com}
\affiliation{Department of Physics and Astronomy, Michigan State University, East Lansing, Michigan 48824, USA}
\affiliation{National Superconducting Cyclotron Laboratory, Michigan State University, East Lansing, Michigan 48824, USA}
\author{D.~P\'erez-Loureiro}
\email{david.perezloureiro@cnl.ca}
\affiliation{National Superconducting Cyclotron Laboratory, Michigan State University, East Lansing, Michigan 48824, USA}
\author{C.~Wrede}
\email{wrede@nscl.msu.edu}
\affiliation{Department of Physics and Astronomy, Michigan State University, East Lansing, Michigan 48824, USA}
\affiliation{National Superconducting Cyclotron Laboratory, Michigan State University, East Lansing, Michigan 48824, USA}
\author{J.~M.~Allen}
\affiliation{Department of Physics, University of Notre Dame, Notre Dame, Indiana 46556, USA}
\author{D.~W.~Bardayan}
\affiliation{Department of Physics, University of Notre Dame, Notre Dame, Indiana 46556, USA}
\author{M.~B.~Bennett}
\affiliation{Department of Physics and Astronomy, Michigan State University, East Lansing, Michigan 48824, USA}
\affiliation{National Superconducting Cyclotron Laboratory, Michigan State University, East Lansing, Michigan 48824, USA}
\author{B.~A.~Brown}
\affiliation{Department of Physics and Astronomy, Michigan State University, East Lansing, Michigan 48824, USA}
\affiliation{National Superconducting Cyclotron Laboratory, Michigan State University, East Lansing, Michigan 48824, USA}
\author{K.~A.~Chipps}
\affiliation{Oak Ridge National Laboratory, Oak Ridge, Tennessee 37831, USA}
\affiliation{Department of Physics and Astronomy, University of Tennesssee, Knoxville, Tennessee 37996, USA}
\author{M.~Febbraro}
\affiliation{Oak Ridge National Laboratory, Oak Ridge, TN 37831, USA}
\affiliation{Department of Physics and Astronomy, University of Tennesssee, Knoxville, Tennessee 37996, USA}
\author{C.~Fry}
\affiliation{Department of Physics and Astronomy, Michigan State University, East Lansing, Michigan 48824, USA}
\affiliation{National Superconducting Cyclotron Laboratory, Michigan State University, East Lansing, Michigan 48824, USA}
\author{M.~R.~Hall}
\affiliation{Department of Physics, University of Notre Dame, Notre Dame, Indiana 46556, USA}
\affiliation{Oak Ridge National Laboratory, Oak Ridge, TN 37831, USA}
\author{O.~Hall}
\affiliation{Department of Physics, University of Notre Dame, Notre Dame, Indiana 46556, USA}
\author{S.~N.~Liddick}
\affiliation{Department of Chemistry, Michigan State University, East Lansing, Michigan 48824, USA}
\affiliation{National Superconducting Cyclotron Laboratory, Michigan State University, East Lansing, Michigan 48824, USA}
\author{A.~Magilligan}
\affiliation{Department of Physics and Astronomy, Michigan State University, East Lansing, Michigan 48824, USA}
\affiliation{National Superconducting Cyclotron Laboratory, Michigan State University, East Lansing, Michigan 48824, USA}
\author{P.~O'Malley}
\affiliation{Department of Physics, University of Notre Dame, Notre Dame, Indiana 46556, USA}
\author{W-J.~Ong}
\affiliation{Department of Physics and Astronomy, Michigan State University, East Lansing, Michigan 48824, USA}
\affiliation{National Superconducting Cyclotron Laboratory, Michigan State University, East Lansing, Michigan 48824, USA}
\author{S.~D.~Pain}
\affiliation{Oak Ridge National Laboratory, Oak Ridge, Tennessee 37831, USA}
\author{S.~B.~Schwartz}
\affiliation{Department of Physics and Astronomy, Michigan State University, East Lansing, Michigan 48824, USA}
\affiliation{National Superconducting Cyclotron Laboratory, Michigan State University, East Lansing, Michigan 48824, USA}
\author{P.~Shidling}
\affiliation{Cyclotron Institute, Texas A \& M University College Station, Texas 77843, USA}
\author{H.~Sims}
\affiliation{University of Surrey, GU2 7XH, Guildford, UK}
\author{P.~Thompson}
\affiliation{Oak Ridge National Laboratory, Oak Ridge, Tennessee 37831, USA}
\affiliation{Department of Physics and Astronomy, University of Tennesssee, Knoxville, Tennessee 37996, USA}
\author{H.~Zhang}
\affiliation{Department of Physics and Astronomy, Michigan State University, East Lansing, Michigan 48824, USA}
\affiliation{National Superconducting Cyclotron Laboratory, Michigan State University, East Lansing, Michigan 48824, USA}
\date{\today}
\begin{abstract}
\noindent \textbf{Background}: Superallowed $0^+ \rightarrow 0^+$ $\beta$ decays are used for precision tests of the standard electro-weak model. The decays of isospin $T=2$ nuclides can be used to test theoretical isospin symmetry breaking corrections applied to extract the CKM matrix element $V_{ud}$ from $T = 0,1$ decays by measuring precise $ft$ values and also to search for scalar currents using the $\beta-\nu$ angular correlation kinematically imprinted in the $\beta$ delayed proton spectrum. Key ingredients include the $Q_{\textrm{EC}}$ value and branching of the superallowed transition and the half life of the parent.
\newline
\newline
\noindent \textbf{Purpose}: To determine a precise experimental $Q_{\textrm{EC}}$ value for the superallowed $0^+ \rightarrow 0^+$ $\beta$ decay of $T=2$ $^{20}$Mg and the intensity of $^{20}$Mg $\beta$-delayed $\gamma$ rays through the isobaric analog state in $^{20}$Na.
\newline
\newline
\noindent \textbf{Method}: A beam of $^{20}$Mg was produced using the in-flight method and implanted into a plastic scintillator surrounded by an array of high-purity germanium detectors used to detect $\beta$-delayed $\gamma$ rays. The high-resolution $\gamma$-ray spectrum was analyzed to measure the $\gamma$-ray energies and intensities.
\newline
\newline
\noindent \textbf{Results}: The intensity of $^{20}$Mg $\beta$-delayed $\gamma$ rays through the isobaric analog state in $^{20}$Na was measured to be $(1.60 \pm 0.04_{\textrm{stat}} \pm 0.15_{\textrm{syst}} \pm 0.15_{\textrm{theo}}) \times 10^{-4}$, where the uncertainties are statistical, systematic, and theoretical, respectively. The $Q_{\textrm{EC}}$ value for the superallowed transition was determined to be $4128.7 \pm 2.2$ keV based on the measured excitation energy of $6498.4 \pm 0.2_{\textrm{stat}} \pm 0.4_{\textrm{syst}}$ keV and literature values for the ground-state masses of $^{20}$Na and $^{20}$Mg.
\newline
\newline
\noindent \textbf{Conclusions}: The $\beta$-delayed $\gamma$-decay branch and $Q_{\textrm{EC}}$ value are now sufficiently precise to match or exceed the sensitivity required for current low-energy tests of the standard model. Future work on the $ft$ value should focus on improving the precision of the $\beta$-delayed proton-decay branch and the half life to match the precision of the present measurements.
\end{abstract}
\maketitle
\section{Introduction}

Superallowed $0^+ \rightarrow 0^+$ $\beta$ decays are used for precision tests of the standard electro-weak model \cite{go19ppnp}. Key nuclear-data ingredients that must be measured experimentally include the $Q_{\textrm{EC}}$ value and branching of the superallowed transition and the half life of the parent. Based on many experiments worldwide over several decades, these quantities have been determined to high precision for the decays of fourteen nuclides with isospin $T=0,1$. This has enabled extraction of the $ft$ value leading to a value of the CKM matrix element $V_{ud}$ following theoretical corrections including those for isospin symmetry breaking (ISB) \cite{ha15prc}. Various nuclear structure methods have been applied to calculate the ISB corrections, which yield different results, and due to this uncertainty it is important to determine which calculations are most reliable \cite{to10prc}. One way to assess this is to make precision measurements of decays for which the ISB corrections are expected to be larger, which may be the case for $T=2$ superallowed decays. By measuring the difference between the raw measured $ft$ value and the corrected $ft$ value expected based on CKM unitarity an empirical value for the ISB correction can be extracted and compared to model predictions. So far, the most precisely measured $T=2$ superallowed decay is that of $^{32}$Ar \cite{bh08prc}. In that case the relative uncertainty of $\Delta ft/ft = 0.77$~\% was marginally sufficient to test ISB models. Future measurements should aim to achieve at least this level of precision and ideally improve upon it.

Another characteristic of $T=2$ decays that can benefit standard-model tests is that the isobaric analog states of the daughters tend to be unbound to proton emission. The $\beta-\nu$ correlation coefficient can be kinematically extracted from  precision measurements of the proton energy spectrum for comparison to the standard-model value and this has been carried out for the case of $^{32}$Ar \cite{ad99prl,ar19arx} to search for scalar-current contributions. A precisely measured $Q_{\textrm{EC}}$ value for the superallowed decay has been prerequisite to an accurate extraction of the $\beta-\nu$ correlation \cite{ad99prl,bl03prl}.

Generally, it would be valuable to improve the data on the decays of other $T=2$ nuclides so that they might be used in addition to $^{32}$Ar \cite{wr10prc}. Indeed, there are major efforts underway to measure the $\beta$-delayed protons for various $T=2$ decays with high precision \cite{sh19hyp}. For $T=2$, $J^{\pi} = 0^+$ $^{20}$Mg, only the ground-state mass excesses $\Delta$ of the parent and daughter are known to sufficient precision. In the present work, we sought to measure the $\beta\gamma$ portion of the superallowed branching to the $T=2$ isobaric analog state (IAS) in $^{20}$Na and its excitation energy $E_{\textrm{x}}$, in order to provide a  precise value for the superallowed $Q_{\textrm{EC}}$.

\section{Experiment}

The experimental procedure has already been described in previous reports \cite{gl15prc,wr17prc,gl18plb,gl19prc}. Briefly, the National Superconducting Cyclotron Laboratory's Coupled Cyclotron Facility accelerated a 170 MeV/u, 60 pnA beam of $^{24}$Mg, which impinged upon a 961 mg/cm$^2$ Be target to produce a fast mixed beam including $^{20}$Mg by projectile fragmentation. The $^{20}$Mg was isolated to a purity of 34 \% using the A1900 fragment separator \cite{mo03nim} and implanted into a plastic scintillator with an intensity of up to 4000 particles per second. The $^{20}$Mg component and contaminant isotones $^{18}$Ne, $^{17}$F, $^{16}$O, and $^{15}$N were identified using standard $\Delta E$-ToF techniques. The scintillator provided signals from the $\beta^+$ decays of implanted ions. $\beta$-delayed $\gamma$ rays were detected using the Segmented Germanium Array (SeGA) of 16 high-purity germanium detectors \cite{mu01nim}.

\section{Analysis}

The present data set and analysis procedures have been detailed in previous reports on the same experiment \cite{gl15prc,wr17prc,gl18plb,gl19prc}. In the present work, we adopt the energy calibration detailed in Ref. \cite{gl15prc} and the efficiency calibration detailed in Ref. \cite{gl18plb}. The energy calibration is based on the well-known room background lines from $^{40}$K and $^{208}$Tl decay at 1460 and 2614 keV, respectively, combined with high-energy $^{20}$Na $\beta$-delayed $\gamma$ rays up to 8.6 MeV. The efficiency calibration is based on a GEANT4 Monte Carlo simulation of the experimental setup that is tuned to reproduce the absolute efficiencies measured in the range of 0.5 to 1.6 MeV using a $^{154}$Eu $\beta$-delayed $\gamma$-ray source positioned on the front face of the scintillator. Coincidences with $\beta$ particles detected in the scintillator were used to remove room background from the SeGA spectra. Coincidences between $\gamma$ rays detected in different SeGA detectors were used to aid in the interpretation of the spectrum and the decay scheme.


\section{Results and Discussion}

\subsection{$Q_{\textrm{EC}}$ value}

As reported in Ref. \cite{gl15prc}, a $\gamma$ ray was observed with a transition energy of $983.73 \pm 0.00_{\textrm{stat}} \pm 0.10_{\textrm{syst}}$ keV corresponding to de-excitation of the well-known 984-keV excited state of $^{20}$Na. A previously unobserved $^{20}$Na $\gamma$ ray was observed, found to be in coincidence with the 984-keV $\gamma$ ray, and measured to have a transition energy of $5514.7 \pm 0.2_{\textrm{stat}} \pm 0.4_{\textrm{syst}}$ keV. By adding these two energies, the excitation energy of the 5514-keV $\gamma$-ray emitting state in $^{20}$Na was determined to be $6498.4 \pm 0.2_{\textrm{stat}} \pm 0.4_{\textrm{syst}}$ keV \cite{gl15prc}. We identified this to be the IAS because of the proximity of its energy to that measured using $\beta$-delayed protons, by the fact that it emitted $\gamma$ rays despite being several MeV above the proton separation energy (the proton width of the IAS is suppressed by isospin conservation) and by the consistency of its main decay branch with shell-model calculations detailed below.

For comparison, the present value for the IAS energy is 27 keV different from, and a factor of 28 more precise than, the previously reported value of $6525 \pm 14$ keV from the most recent data evaluation \cite{ma14npa}, which was based on several measurements of $^{20}$Mg $\beta$-delayed proton emission \cite{mo79prl,go92prc,pi95npa}; it is consistent with the proton-based values of $6496 \pm 3$ keV \cite{lu16epj} and $6523 \pm 28$ keV \cite{su17prc} reported after Ref. \cite{gl15prc}. Presently, we combine our value with the evaluated literature values \cite{wa17cpc} for the $^{20}$Na \cite{wr10prc} and $^{20}$Mg \cite{ga14prl} mass excesses and report a new $Q_{\textrm{EC}}$ value of $4128.7 \pm 2.2$ keV for the superallowed $0^+ \rightarrow 0^+$ $\beta$ decay of $^{20}$Mg. This value is of comparable precision to the superallowed $Q_{\textrm{EC}}$ value of $^{32}$Ar \cite{ma14npa,wa17cpc}. The uncertainty is now dominated by the uncertainties associated with the ground-state mass excesses of $^{20}$Mg ($\pm 1.9$ keV) and $^{20}$Na ($\pm 1.1$ keV) \cite{wa17cpc}.

\subsection{Superallowed branching}

As reported in Ref. \cite{gl15prc}, the only observed primary $\gamma$-ray transition originating from the IAS in $^{20}$Na was to the 984-keV final state. The intensity of this $\gamma$-ray transition has now been determined to be $I_{\beta\gamma} = (1.45 \pm 0.04_{\textrm{stat}} \pm 0.15_{\textrm{syst}}) \times 10^{-4}$ by integrating the peak above background and applying the efficiency, where the total number of decays was determined using the 984-keV peak. The total uncertainty is dominated by systematic effects associated with the efficiency \cite{gl18plb,wr17prc}.

Indeed, the shell-model calculations detailed below predict that this should be the dominant $\gamma$-ray transition from the IAS. However, the shell model also predicts that $\approx 10$\% of the $\gamma$-ray branching should be fragmented across transitions to other states. Since we were not sensitive to the other weak $\gamma$-ray transitions, we apply a correction of $+(10\pm10)$\% to our measured value to determine a total intensity of $^{20}$Mg $\beta$-delayed $\gamma$ rays through the IAS of $I_{\beta\gamma} = (1.60 \pm 0.04_{\textrm{stat}} \pm 0.15_{\textrm{syst}} \pm 0.15_{\textrm{theo}}) \times 10^{-4}$. This value constitutes the first reported measurement of the $^{20}$Mg $\beta$-delayed $\gamma$-decay intensity through the IAS.

The present value is more than two orders of magnitude lower than the intensity of $^{20}$Mg $\beta$-delayed protons through the same state, which has been reported to be $I_{\beta p} = (3.3 \pm 0.4) \times 10^{-2}$ \cite{pi95npa} and $I_{\beta p} = (2.2 \pm 0.2) \times 10^{-2}$ \cite{lu16epj} in the most precise previous works; the weighted average of all published values is $I_{\beta p} = (2.54 \pm 0.17) \times 10^{-2}$ \cite{go92prc,pi95npa,lu16epj,su17prc}. Adding $I_{\beta\gamma}$ and $I_{\beta p}$ yields the total branching $I_{\beta} = (2.55 \pm 0.17) \times 10^{-2}$ for the superallowed $0^+ \rightarrow 0^+$ $\beta$ decay of $^{20}$Mg, which is now evidently dominated by $I_{\beta p}$ and the associated uncertainty.

Regarding the proton branch, we have recently reported indirect evidence for a previously unobserved $2.70 \pm 0.23$ MeV $\beta$-delayed proton transition through the IAS with an intensity of $(2.12 \pm 0.07) \times 10^{-3}$ based on the Doppler broadening of $^{19}$Ne $\gamma$ rays \cite{gl19prc}. However, the uncertainty associated with the energy of this proton branch is too large to confirm that it is indeed from the IAS instead of another nearby state. Measuring these protons directly could provide confirmation. For example, using a proton-$\gamma$-ray coincidence method might be an effective way to distinguish these protons from proton emissions following Gamow-Teller transitions.

\subsection{$ft$ value}

\begin{figure}
\includegraphics[width=0.5\textwidth]{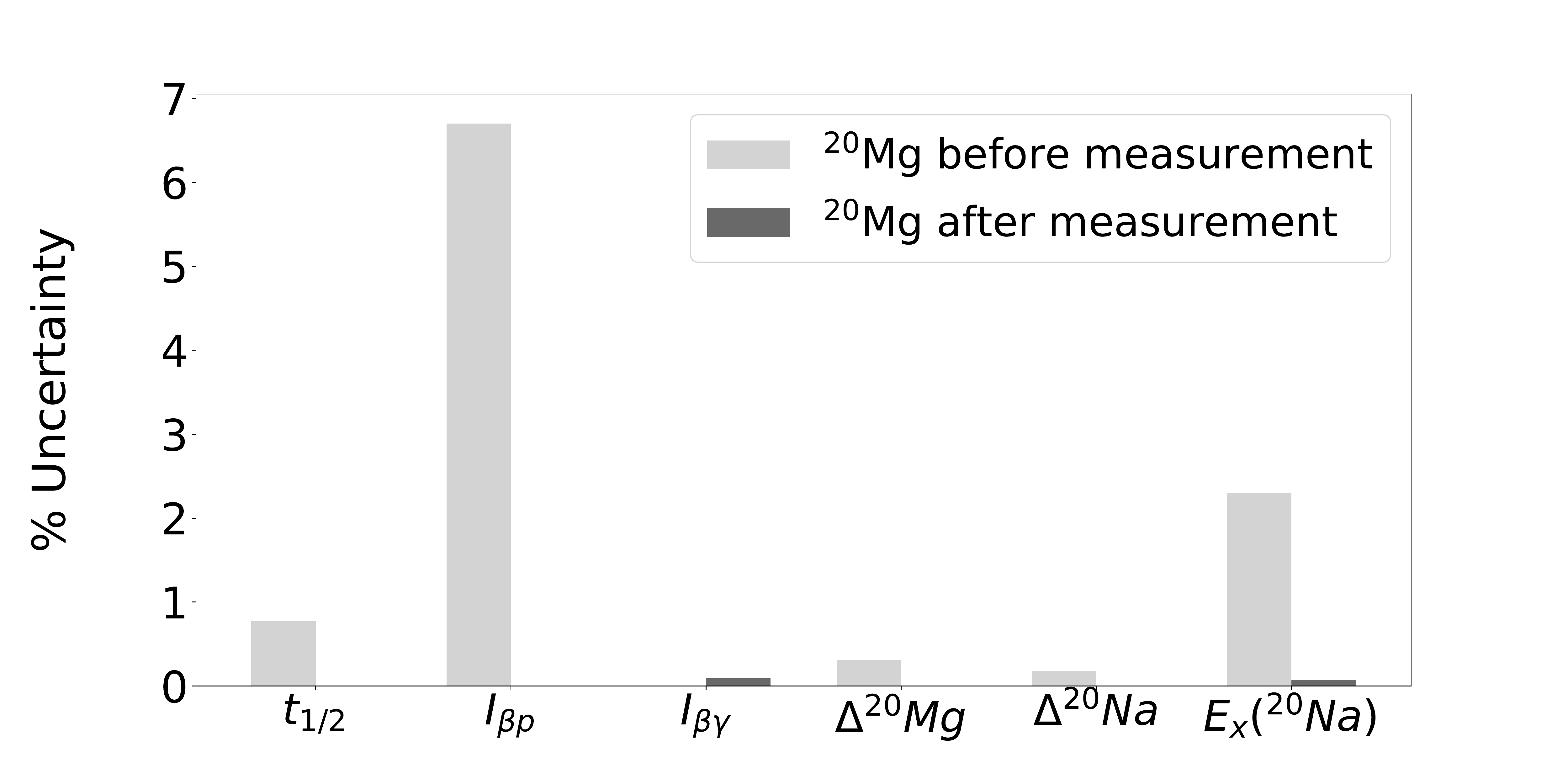}
\caption{Uncertainties associated with the $ft$ value for the superallowed $0^+ \rightarrow 0^+$ $\beta$ decay of $^{20}$Mg. Light grey denotes the uncertainties before the present measurements. Dark grey denotes the uncertainties for quantities measured in the present experiment. The ``before'' value for $I_{\beta\gamma}$ is omitted because it was not measured previously.}
\label{fig: Unc}
\end{figure}

Figure~\ref{fig: Unc} shows contributions to the uncertainty associated with the $ft$ value for the superallowed $0^+ \rightarrow 0^+$ $\beta$ decay of $^{20}$Mg, before and after the present measurement. The uncertainty associated with the half life $t_{1/2}$ was adopted from the TUNL Nuclear Data Evaluation Project \cite{tunl}, which is strongly influenced by two recent measurements \cite{lu16epj,su17prc}. The uncertainties associated with the weighed average $I_{\beta p}$ value, the mass-excess literature values, and the energy of the IAS discussed above were adopted. The effects of the mass excesses and IAS energy were incorporated by propagating their associated uncertainties in the calculation of the phase-space factor, \emph{f}.

Both the excitation energy and the $\beta$-delayed $\gamma$-decay branch determined in the present experiment exceed the precision and accuracy required to benchmark ISB corrections to the \emph{ft} values for superallowed decays. Future experiments should focus on improving the precision of the $\beta$-delayed proton branch by at least an order of magnitude. Doing so will require higher precision measurements of known protons and also higher sensitivity to detect potentially missing branches such as the 2.7 MeV protons discussed above. In the case of $^{32}$Ar, two complementary techniques were employed to achieve the required precision \cite{ad99prl,bh08prc}. It would also be beneficial to improve the precision of the half life by at least a factor or two.

\subsection{Shell-model calculations}

For comparison with the experimental values, we have performed shell-model calculations using the USDB-cdpn interaction \cite{ri12prc}. These calculations predict the IAS at an excitation energy of 6.84 MeV, which is within 0.34 MeV of the experimental value from the present work. The ratio $I_{\beta\gamma}$/$I_{\beta p}$ is equal to the ratio $\Gamma_{\gamma}/\Gamma_p$, where $\Gamma_{\gamma}$ and $\Gamma_p$ are the partial widths for $\gamma$-ray emission and proton emission from the IAS, respectively. Based on the present value of $I_{\beta\gamma}$ and the average of the literature values of $I_{\beta p}$, we obtain the experimental value of $\Gamma_{\gamma}/\Gamma_p = 0.0063 \pm 0.0009$ for the IAS. The shell-model calculations yield $\Gamma_{\gamma}=7.5$~eV and $\Gamma_p = 580$~eV. The latter value was based on the shell-model value $C^2S = 1.49 \times 10^{-4}$ of the spectroscopic factor and a single-particle width of 0.70 MeV for proton emission to the first excited state of $^{19}$Ne, and a scaling of the widths for proton branches to other states using the experimental branches reported in the literature \cite{pi95npa}. The transition to the first excited state was used as the anchor for the theoretical prediction of $\Gamma_p$ because it had the smallest theoretical uncertainties. Ref. \cite{pi95npa} was chosen for the scaling because it was consistent with the Doppler broadening analysis in Ref. \cite{gl19prc}. The resulting theoretical value of $\Gamma_{\gamma}/\Gamma_p = 0.013$ is a factor of two larger than the experimental one, which is good agreement considering the theoretical uncertainties, which can be assessed by employing different interactions.

To this end, shell-model calculations were also carried out with two new USD-type interactions, USDC and USDI, that are in preparation for publication \cite{ma19tbs}. These interactions were developed to directly incorporate isospin symmetry breaking. USDC is based the same RGSD Hamiltonian \cite{hj95pr} that was used for USDB. RGSD is an \emph{ab initio} Hamiltonian developed using theoretical methods to calculate the effective two body matrix elements with a renormalized G-matrix starting from NN interactions. USDI is based on IMSRG Hamiltonians \cite{st17prl,st19arx}. The new interactions predict the excitation energy of the IAS to be 6.722 MeV (USDC) and 6.729 MeV (USDI), placing them about 0.1 MeV closer to experiment than USDB-cdpn. The $C^2S$ value for the first excited state in $^{19}$Ne is higher in these interactions than it is in USDB-cdpn, which results in $\Gamma_p$ values of 1026 eV (USDC) and 1058 eV (USDI) when the same scaling is used as for the USDB-cdpn calculation above. Along with a small decrease in $\Gamma_{\gamma}$, the $\Gamma_{\gamma}/\Gamma_p$ ratios are 0.0066 (USDC) and 0.0065 (USDI), which are in agreement with the experiment.

Despite the fact that the proton emission from the $T=2$ IAS to $T=1/2$ states is isospin forbidden, the proton emission dominates over $\gamma$ decay. The shell-model calculations demonstrate that this is not due to a large isospin impurity of the IAS, but rather due to its energy far above the proton-emission threshold and the corresponding lack of suppression by the Coulomb barrier. Therefore, the large isospin-forbidden proton branch is consistent with the good quadratic isobaric multiplet mass equation description of the $A=20, T=2$ quintent \cite{gl15prc}.

\section{Conclusions}

We have measured the intensity of $^{20}$Mg $\beta$ delayed $\gamma$ rays through the isobaric analog state in $^{20}$Na for the first time. We also report a new precision $Q_{\textrm{EC}}$ value for the superallowed $0^+ \rightarrow 0^+$ $\beta$-decay transition. Finally, we point out indirect evidence for a new proton emission branch from the $^{20}$Mg IAS in $^{20}$Na. These quantities are essential ingredients in the calculation of the $ft$ value for the $T=2$ superallowed transition, which can be used to test theoretical calculations of isospin symmetry breaking needed to extract the CKM matrix element $V_{ud}$ from $T=0,1$ superallowed transitions. The $Q_{\textrm{EC}}$ value is an essential input for precision measurements of the $\beta-\nu$ correlation coefficient, which can be used to search for scalar currents.

\section{ACKNOWLEDGMENTS}

We gratefully acknowledge the NSCL staff for technical assistance and for providing the $^{20}$Mg beam and to Lijie Sun for providing comments. This work was supported by the National Science Foundation (USA) under Grants No. PHY-1102511, No. PHY-1419765, No. PHY-1404442, No. PHY-1430152, No. PHY-1565546, No. PHY-1913554, and No. PHY-1811855, the US Department of Energy, Office of Science, under Award No. DE-SC0016052, Contract No. DE-AC05-00OR22725, and the US Department of Energy National Nuclear Security Administration under Awards No. DE-NA0003221 and No. DE-NA0000979.

\end{document}